\documentclass[aps,rmp,floats,floatfix,twocolumn,superscriptaddress,final]{revtex4-2}

\usepackage{float}
\usepackage{amsmath,bm}
\usepackage{graphicx}
\usepackage{ragged2e}
\usepackage{xcolor}
\usepackage{lipsum}
\usepackage[percent]{overpic}
\usepackage[caption=false]{subfig}

\usepackage[hidelinks]{hyperref}
\hypersetup{colorlinks=false,
    linkcolor=black,
    citecolor=black,
    urlcolor=black,
    pdftitle={Mapping change in higher-order networks with multilevel and overlapping communities},
    pdfauthor={Anton Holmgren, Daniel Edler, Martin Rosvall}
}
\usepackage{xurl}
\usepackage{orcidlink}

\usepackage[capitalise]{cleveref}

\makeatletter
\newcommand*{\balancecolsandclearpage}{%
  \close@column@grid
  \cleardoublepage
  \twocolumngrid
}
\makeatother

\usepackage{listings}
\crefname{lstlisting}{listing}{listings}
\Crefname{lstlisting}{Listing}{Listings}
\setcitestyle{numbers,super,comma,sort&compress}

\newcommand{\CC}{C\nolinebreak\hspace{-.05em}\raisebox{.4ex}{\tiny\bf +}\nolinebreak\hspace{-.10em}\raisebox{.4ex}{\tiny\bf +}}

\begin{document}

\title{Mapping change in higher-order networks\\ with multilevel and overlapping communities}

\author{Anton Holmgren \orcidlink{0000-0001-5859-4073}}
\thanks{Corresponding author}
\email{anton.holmgren@umu.se}
\affiliation{Integrated Science Lab, Department of Physics, Ume{\aa} University, SE-901 87 Ume{\aa}, Sweden}

\author{Daniel Edler \orcidlink{0000-0001-5420-0591}}
\affiliation{Integrated Science Lab, Department of Physics, Ume{\aa} University, SE-901 87 Ume{\aa}, Sweden}
\affiliation{Gothenburg Global Biodiversity Centre, Department of Biological and Environmental Sciences, University of Gothenburg, SE-405 30 Gothenburg, Sweden}

\author{Martin Rosvall \orcidlink{0000-0002-7181-9940}}
\email{martin.rosvall@umu.se}
\affiliation{Integrated Science Lab, Department of Physics, Ume{\aa} University, SE-901 87 Ume{\aa}, Sweden}

\date{\today}

\begin{abstract}
New network models of complex systems use layers, state nodes, or hyperedges to capture higher-order interactions and dynamics.
Simplifying how the higher-order networks change over time or depending on the network model would be easy with alluvial diagrams, which visualize community splits and merges between networks.
However, alluvial diagrams were developed for networks with regular nodes assigned to non-overlapping flat communities.
How should they be defined for nodes in layers, state nodes, or hyperedges?
How can they depict multilevel, overlapping communities?
Here we generalize alluvial diagrams to map change in higher-order networks and provide an interactive tool for anyone to generate alluvial diagrams.
We use the alluvial generator to illustrate the effect of modeling network flows with memory in a citation network, distinguishing multidisciplinary from field-specific journals.
\end{abstract}

\maketitle


\section*{Introduction}

Complex systems are inherently dynamic.
Their components influence each other through various informational and physical processes, changing interaction patterns over time.
Researchers represent these interactions with networks \cite{edler2017infomap,calatayud2020positive,farage2021identifying,joaquin2021regularities,neuman2022pisa,edler2022infomap,rojas2022natural}
and simplify their organization with community-detection algorithms \cite{rosvall2008maps,fortunato2010community,schaub2017many,traag2019louvain,peixoto2019bayesian}.
For example, community-detection algorithms that model the various processes as flows on networks assign nodes to possibly nested modules of typically densely connected nodes, among which the network flows persist relatively long \cite{rosvall2008maps}. 
Identifying modules in multiple networks with shared nodes enables exploring organizational changes when the systems they represent change over time or between states: Modules merge and split when groups in students' social networks form and dissolve during school days, or new research fields emerge when old fields fuse or break and move apart. 
Various summary statistics can quantify these structural changes \cite{danon2005comparing,amelio2017correction,newman2020improved}, but they destroy essential information about how the networks change.

Alluvial diagrams with modules represented as stacks of blocks joined by stream fields were introduced to reveal network organizational changes by depicting merging and splitting modules \cite{rosvall2010mapping}.
Researchers have successfully used them to map shifting regional tendencies in urban networks \cite{liu2013featured}, study dynamics of hot topics in research fields \cite{ruan2017detecting,pal2022predicting}, track changing bitcoin user activity \cite{remy2017tracking}, and explore evolving media channel preferences across crisis phases \cite{petrun2021disasters}.
Generating alluvial diagrams requires dedicated software to remove tedious manual work. 
However, current applications to generate alluvial diagrams work only for standard networks partitioned into modules.

Today researchers use temporal, multilayer, and memory networks to capture interactions in complex systems with higher accuracy \cite{kivela2014multilayer,rosvall2014memory,de2015identifying,de2016physics,xu2016representing,lambiotte2019networks} and multilevel modular solutions to reveal more regularities in their organization \cite{rosvall2011multilevel,peixoto2014hierarchical}.
Multilayer networks can represent networks over time with links in time-windowed layers.
Memory networks can represent higher-order network flow models where the transition rates depend on the current node and previously visited nodes. Both representations enable overlapping modules.
Mapping change in these rich network representations requires generalizing alluvial diagrams and their generators to higher-order networks with multilevel and overlapping modular solutions.

\begin{figure*}[htp!]
    \centering
    \subfloat{\label{fig:modular-schematic-a}}
    \subfloat{\label{fig:modular-schematic-b}}
    \subfloat{\label{fig:modular-schematic-c}}
    \subfloat{\label{fig:modular-schematic-d}}
    \subfloat{\label{fig:modular-schematic-e}}
    \vspace*{12pt}
    \begin{overpic}[width=\linewidth]{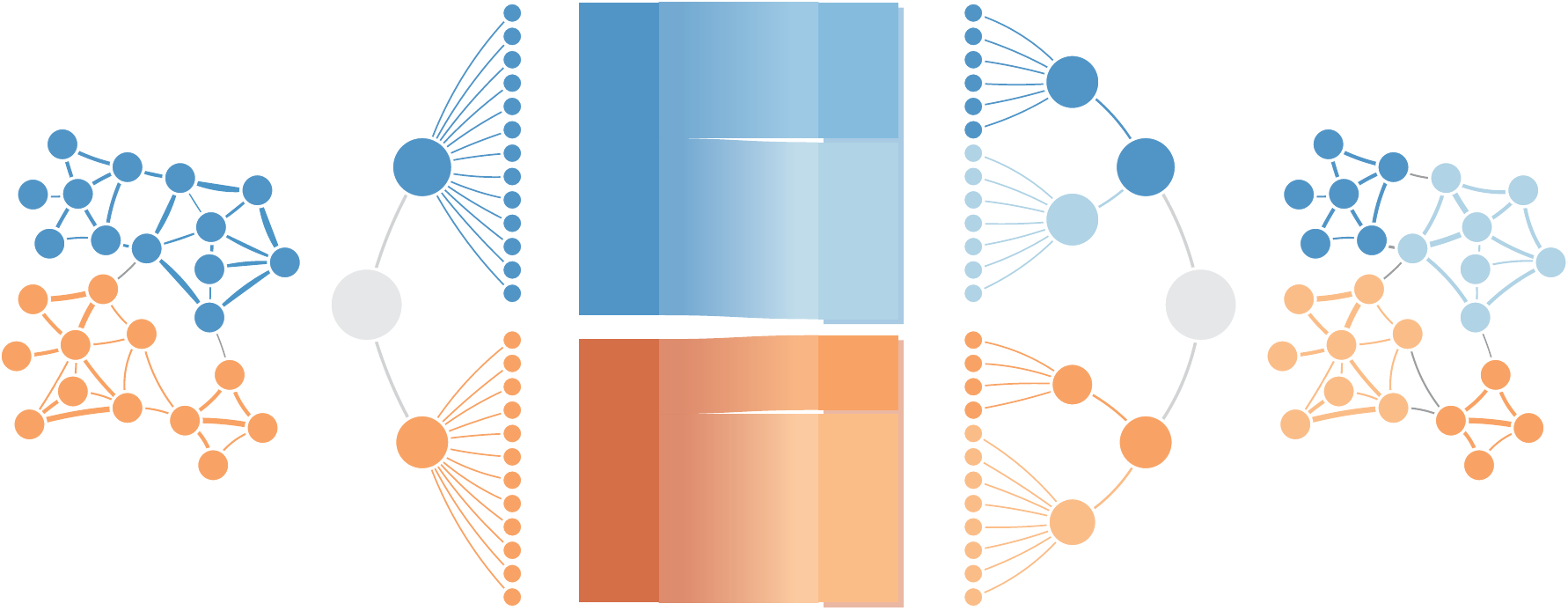}
        \put(0,41){\small\sffamily{\textbf{a}\hspace{2pt} Modular network}}
        \put(18,41){\small\sffamily{\textbf{b}\hspace{2pt} Two-level solution}}
        \put(39.5,41){\small\sffamily{\textbf{c}\hspace{2pt} Alluvial diagram}}
        \put(61,41){\small\sffamily{\textbf{d}\hspace{2pt} Multilevel solution}}
        \put(82,41){\small\sffamily{\textbf{e}\hspace{2pt} Multilevel network}}
        
        \put(18,-3){\small\sffamily Level}
        \put(26.4,-3){\small $1$}
        \put(32.3,-3){\small $2$}
        \put(61.6,-3){\small $3$}
        \put(68,-3){\small $2$}
        \put(73,-3){\small $1$}
    \end{overpic}
    \vspace*{12pt}
    \caption{Schematic alluvial diagram of a multilevel network structure.
    a) A weighted network with modular structure, organized into a two-level solution in b).
    c) An alluvial diagram representation of the solutions in panels (b) and (d) using the same colors. Columns of blocks represent modules with heights proportional to the contained flow volume. The leftmost column is an ordinary two-level alluvial diagram representation. The multilevel representation to the right shows multiple levels, with the background showing the top-level organization. Stream fields connect modules in the left and right columns that share nodes.
    d) Multilevel solution of the network in (e).}
    \label{fig:modular-schematic}
\end{figure*}
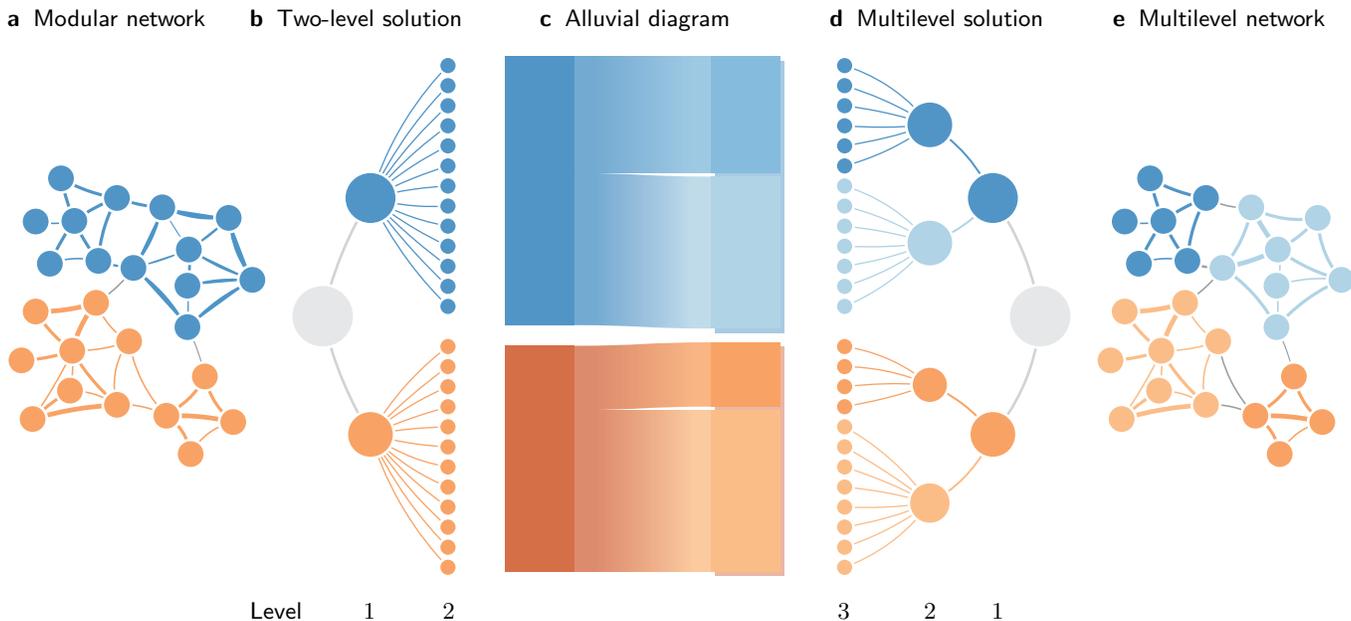

Here we introduce alluvial diagrams for multilayer and memory networks with multilevel and overlapping modular solutions.
We demonstrate a new alluvial generator for higher-order networks available for anyone to use at \url{https://www.mapequation.org/alluvial}\cite{mapequation2022alluvial}, and illustrate how we use it in three case studies revealing: 
significant changes in the multilevel organization of science over six years using parametric bootstrap resampling, multidisciplinary journals in a second-order network representation of citation flows, and the effects of multilayer representation of a collaboration hypergraph.

\section*{Methods}

Alluvial diagrams depict changes in the modular composition between networks with stacks of blocks representing the modules (\cref{fig:modular-schematic}).
Each block's height is proportional to the flow volume of the corresponding module -- the total visit probability of all nodes in the module.
To highlight structural change between multiple networks, a vertical stack of blocks represent each network's modular structure, and horizontal stream fields connect blocks that share nodes across neighboring networks.
Like block heights, stream-field heights are proportional to the flow volume of the node overlap between corresponding modules.
To reduce clutter, we order stream fields to minimize their overlap.

We use Infomap to search for multilevel modular structures with nested submodules.\cite{mapequation2022software,rosvall2011multilevel}
Infomap optimizes the map equation, the average per-step codelength on a modular description of a random walk modeling network flows.\cite{rosvall2008maps}
The modules are groups of nodes where the random walker spends a relatively long time compared to exiting it and entering other modules.
While we focus on modules derived from Infomap, alluvial diagrams work with output from any community detection or hierarchical data clustering method.

\subsection*{Mapping change in networks with multilevel communities}

We extend alluvial diagrams to multilevel network partitions by nesting submodules in super-modules with adaptive module distances.
The right multilayer stack of the schematic alluvial diagram in \cref{fig:modular-schematic-c} illustrates. 
In the spirit of cartography, we put blocks corresponding to the top-level modules in a bottom layer to highlight the large-scale organization and provide a cleaner visualization.
Optionally, we display finer-level structures in layers above the bottom layer.
The right multilayer stack in \cref{fig:modular-schematic-c} expands the left stack's single layer with one such extra layer corresponding to the four submodules of the multilevel modular solution. 
To show that deeper submodules are more closely related than their larger parent modules, we draw sibling submodules closer together than other modules.
Specifically, we halve the distance between two adjacent modules for each level down in the multilevel solution.

\subsubsection*{Multilevel significance clustering}
\label{sec:sigclu}

To separate trends from mere noise in the module assignments, we extend the significance clustering method described in ref.~\onlinecite{rosvall2010mapping} to multilevel partitions.
The approach has three main steps:
First, we search for optimal multilevel partitions for each network using Infomap.
Then, to assess these partitions' robustness to slight perturbations in the data, we create a large number of independent bootstrap networks.
For each bootstrap network, we search for the optimal multilevel partition using Infomap as for the original network.
Finally, we summarize the variability in the bootstrap partitions by applying the significance clustering method introduced in ref.~\onlinecite{rosvall2010mapping} extended to multilevel partitions.
For each level in the multilevel solution of the original network, we search for the largest subset of nodes in each module or submodule that are also clustered together in at least a fraction~$p$ of solutions obtained from the parametric bootstrap procedure.

Searching for significant subsets in multilevel solutions is computationally more demanding than for ordinary two-level partitions.
To improve the performance, we trivially parallelize the algorithm by running each module or submodule in separate threads.

\subsection*{Mapping change in higher-order networks}

We generalize alluvial diagrams to multilayer and memory networks.
Multilayer networks can model different modes of interaction or interactions that change over time in different layers.
Memory networks can model dynamics that depend on from where the flows come.
Infomap represents both higher-order networks with so-called \emph{state nodes}.\cite{edler2017mapping}
In higher-order networks, we call ordinary nodes \emph{physical nodes} to distinguish them from state nodes.
In a multilayer network, one state node for each physical node and layer represents the physical node in the layer.\cite{de2015identifying}
In a second-order memory network with memory of the previous step, one state node for each physical node and incoming link represents the physical node for flows incoming along that link.\cite{rosvall2014memory}
Physical nodes with multiple state nodes and different outgoing links can model higher-order dynamics on the network.

\begin{figure}[htp!]
    \centering
    \subfloat{\label{fig:modular-2nd-order-a}}
    \subfloat{\label{fig:modular-2nd-order-b}}
    \subfloat{\label{fig:modular-2nd-order-c}}
    \vspace*{12pt}
    \begin{overpic}[width=\linewidth]{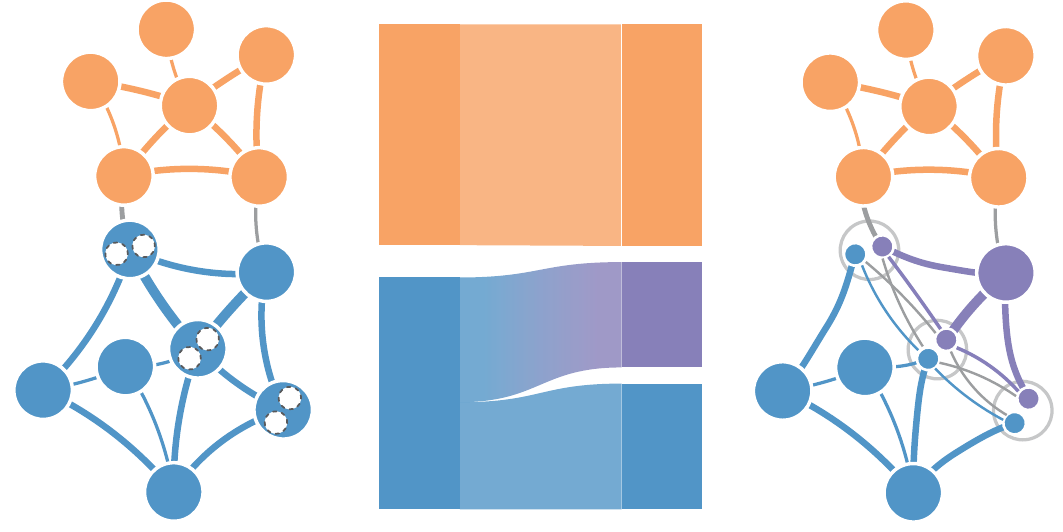}
        \put(5,51){\small\sffamily{\textbf{a}}}
        \put(35,51){\small\sffamily{\textbf{b}}}
        \put(73,51){\small\sffamily{\textbf{c}}}
    \end{overpic}
    \caption{Schematic first-order and higher-order networks and alluvial diagram representation.
    a) A first-order network with pseudo-states (white dashed circles) matching the state nodes in (c).
    b) An alluvial diagram representation.
    c) A higher-order network with state nodes (small blue and purple circles) in the physical nodes. We only show the state nodes whose physical nodes are present in two modules.}
    \label{fig:modular-2nd-order}
\end{figure}
In theory, using alluvial diagrams for higher-order networks is no different than for ordinary networks.
In practice, the many possible combinations of first- and higher-order networks, memory networks with different memory, and multilayer networks with different layers make it challenging to determine node equality in different networks because we need to match nodes across networks to draw stream fields between modules.
While alluvial diagrams require networks to share a significant fraction of physical nodes, we also need their state nodes to match since they are the smallest components of higher-order networks.
With no universal solution to this node-matching problem, we discuss some challenges and how we choose to solve them.

\subsubsection*{First- and higher-order networks}

Alluvial diagrams with first- and higher-order networks require matching different node types:
First-order networks have only physical nodes, but higher-order networks have physical nodes and state nodes.
We illustrate this schematically in \cref{fig:modular-2nd-order} with a first-order network in \cref{fig:modular-2nd-order-a} and a higher-order network with state nodes as smaller circles inside the physical nodes in \cref{fig:modular-2nd-order-c}.
We consider only hard module boundaries in the first-order network, whereas modules overlap in the higher-order network when
physical nodes' state nodes are assigned to different modules.
In \cref{fig:modular-2nd-order-c}, the modules overlap in the physical nodes containing the purple and blue state nodes.

As we need a one-to-one match across networks to draw stream fields, we cannot match all state nodes in the higher-order network to one first-order node.
To overcome this problem, we first split the first-order nodes into pseudo-state nodes, which we depict with small dashed circles in \cref{fig:modular-2nd-order-a}.
We create as many pseudo-state nodes as there are state nodes in the matching physical node in the higher-order network.
Then, we divide first-order node $i$'s flow volume $\pi_i^{(1)}$ among its pseudo-states~$\alpha$ proportionally to their matching state nodes' fraction of the flow $\pi_{i\alpha}^{(2)}$ as
\begin{equation}
    \pi_{i\alpha}^{(1)} = \pi_i^{(1)} \frac{\pi_{i\alpha}^{(2)}}{\pi_i^{(2)}}.
\end{equation}
This procedure gives a one-to-one match between nodes in first- and higher-order networks, and we can draw multiple stream fields from a single first-order node (\cref{fig:modular-2nd-order-b}).

\subsubsection*{Memory networks}

Drawing alluvial diagrams for memory networks requires matching state nodes representing corresponding memory in different networks.
We match state nodes across networks by encoding their memory in their ids such that state nodes representing the same memory share the same id in different networks.
As long as the networks are not too large, we can encode memory of order $n$ into a single binary number by dividing the binary number into $n$ parts:
We divide the number into two parts in a second-order memory network with memory of the previous step.
With $N$~physical nodes, we use the $b = \lceil \log_2 N \rceil$ most significant bits of the state id to encode the previously visited node $i$ and the $b$ least significant bits to encode the currently visited node $j$, resulting in the state id
\begin{equation}
    \alpha_{i \to j} = i \ll b + 1 \, \lor \, j,
\end{equation}
where $\ll$ is the arithmetic left-shift operator and $\lor$ is the logical or.
For example, we encode the link from physical node 2 to physical node 3 along the path represented by the trigram $1\to 2\to 3$ as
\begin{align}
    \alpha_{1 \to 2} &= 1 \ll 2 + 1 \lor 2 = 1000_2 \lor 10_2 = 1010_2 = 10, \nonumber \\
    \alpha_{2 \to 3} &= 2 \ll 2 + 1 \lor 3 = 10000_2 \lor 11_2 = 10011_2 = 19, \nonumber
\end{align}
resulting in the directed link $10 \to 19$ between state nodes 10 and 19.
This encoding scheme works for up to $N = 2^{16} = 65,536$ physical nodes with 32-bit ids and second-order memory.

\subsubsection*{Multilayer networks}

When comparing multiple multilayer networks with $N$ layers, we encode the physical node $i$ in layer $l$ with id
\begin{equation}
    \alpha_{i,l} = i \ll b + 1 \, \lor \, l,
\end{equation}
where $N$ is the largest layer id represented with $b = \lceil \log_2 N \rceil$ bits.
For multilayer networks, this encoding scheme is available in Infomap using the flag \texttt{-{}-matchable-multilayer-ids N}.

Alluvial diagrams can also visualize the layers of multilayer networks, each as a separate network.
In this case, node matching is trivial as physical nodes are unique in each layer.
The stream fields then connect modules that span layers.

\subsection*{Alluvial diagram generator}

We have implemented an interactive web application that generates alluvial diagrams, available for anyone to use at {\small\url{https://www.mapequation.org/alluvial}}.
We implemented it as a client-side web application to enable researchers to use our application without programming experience or those working with sensitive data.
All code runs locally in the user's web browser, and the web application does not store or upload network data to any server.
We implemented it using TypeScript and React, and we display the diagrams using scalable vector graphics (SVG) (see Fig.~S2 in the SI for how we model the data structures).

\begin{figure}[htp!]
    \centering
    \includegraphics[width=\linewidth]{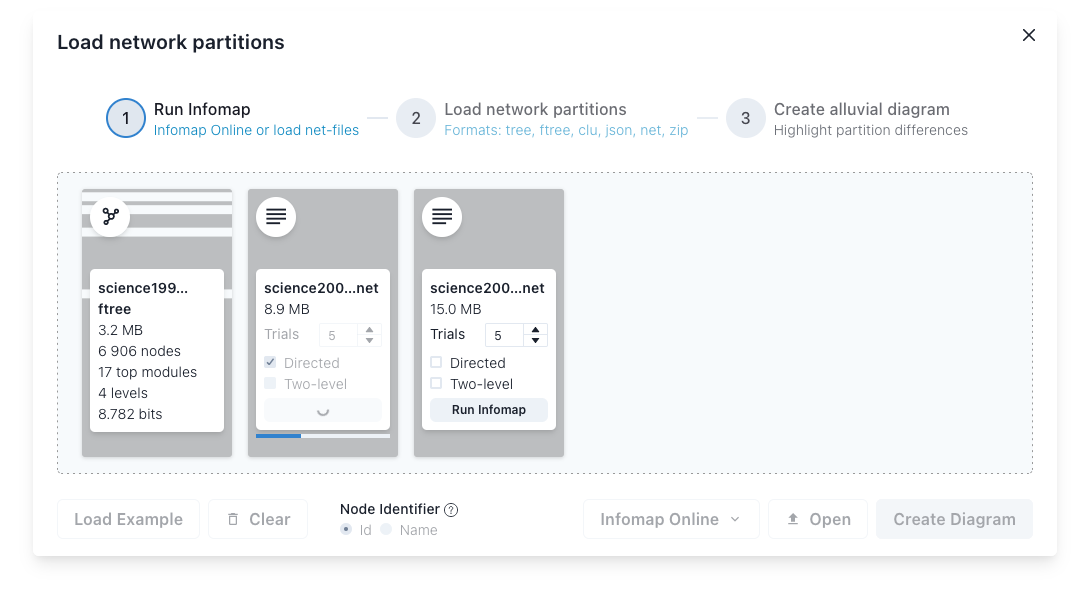}
    \caption{Loading networks in the alluvial diagram generator. Three networks are loaded in different stages, shown as gray rectangles. The leftmost network has communities detected by Infomap. Infomap runs on the second network and has completed two out of five optimization trials. Infomap has yet to start identifying communities in the rightmost network. When all have finished, or when loading networks with communities from \CC\ Infomap, the user can select ``Create Diagram'' to create an alluvial diagram.}
    \label{fig:load-networks}
\end{figure}

While the most efficient community detection pipeline is to run the stand-alone \CC\ version of Infomap and load the resulting partitions, we have embedded a version of Infomap compiled to JavaScript with Emscripten.\cite{zakai2011emscripten}
This embedded Infomap version supports the same network inputs as \CC\ Infomap, but only a subset of Infomap's features, including reading directed or undirected input, choosing the number of optimization trials, and searching for multilevel or two-level solutions (\cref{fig:load-networks}).
We defer the specification of input formats to Sec.~SI.1.
We also support loading solutions from Infomap Online,\cite{mapequation2022infomaponline} a fully featured web-based version of Infomap.

With loaded networks, the interface shows the user a top-level view of the alluvial diagram (Fig.~S1). The user can manipulate the diagram in several ways: expand modules to reveal their submodules, reorganize networks and modules for clarity, highlight modules or individual nodes with different colors, and change the diagram width and height.
While we have implemented the features and use cases we think most researchers use, we can imagine feature requests for specific use cases.
By supporting export to SVG, researchers can modify the diagrams to their needs in any vector graphics application.

\begin{figure*}[htp!]
    \centering
    \includegraphics[width=0.9\textwidth]{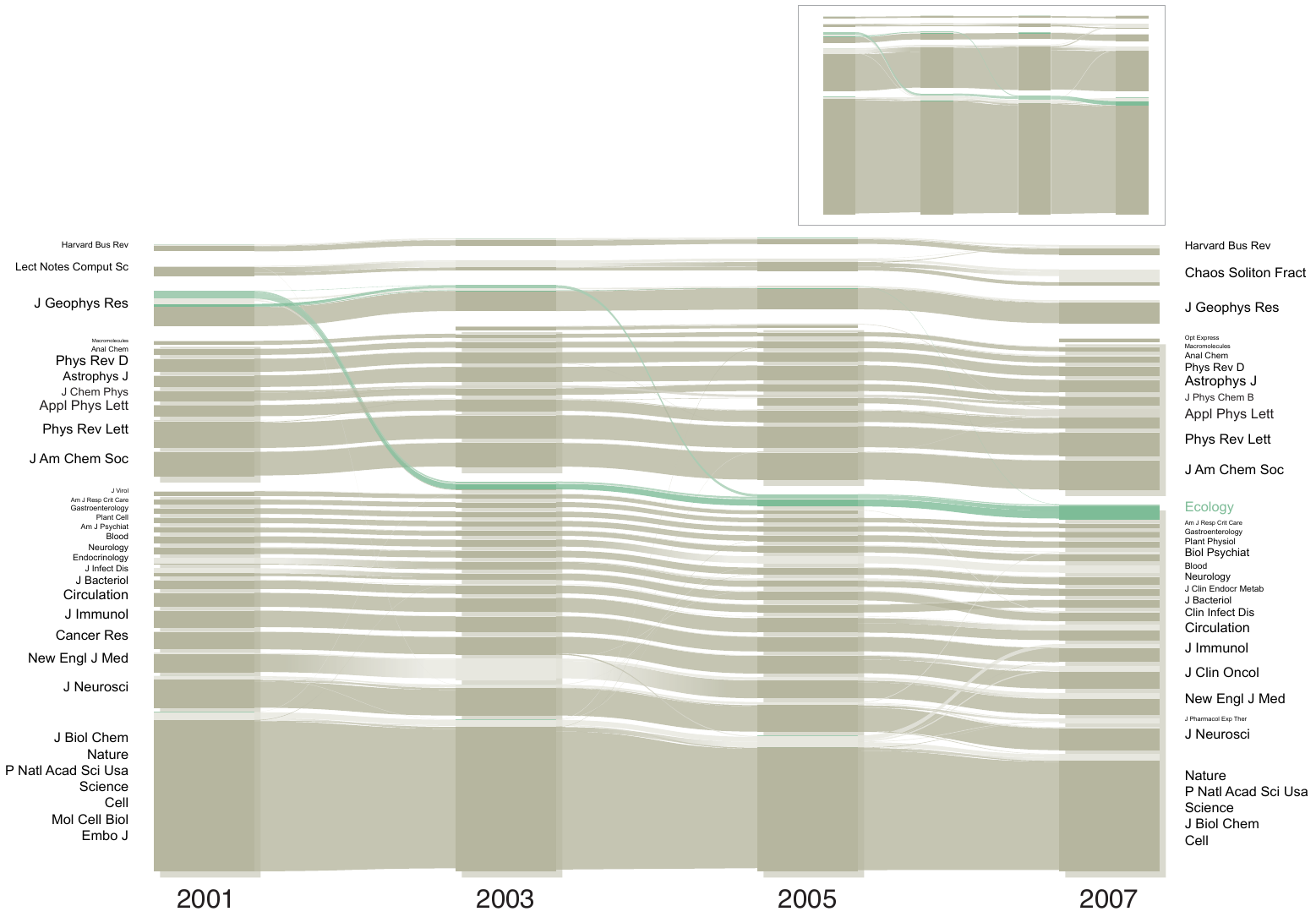}
    \caption{Multilevel organization of science, 2001--2007. The journals organize into five large research areas, further divided into research fields. We show the top-level organization of earth sciences, economics, and computer science (the three small modules at the top). We show the finer division into research fields for the physical and life sciences. We sort the fields or areas by their citation flow and show the top-ranking journal in each research field or area.
    Finally, we highlight the insignificant assignment of journals clustered with Ecology in 2001 to a significantly distinct research field since 2003.
    At the top level shown in the inset, the same journals cluster significantly with the life sciences since 2007.}
    \label{fig:science-alluvial}
\end{figure*}

\section*{Results}

We highlight different visualization challenges in three case studies using multilevel, higher-order, and multilayer networks.
In all cases, we use Infomap to identify optimal multilevel solutions using unrecorded teleportation to links with minimal impact from the teleportation rate on the results.\cite{lambiotte2012ranking}

\subsection*{Robust multilevel citation networks}

First, we highlight the multilevel organization of science into research areas and fields.
We use data from Thomson-Reuters Journal Citation Reports.\cite{rosvall2010mapping}
The data include citations between journals published from 2001--2007, divided into four two-year periods.
The networks have, on average, $7,490$ nodes representing journals and $586,295$ integer-weighted links representing the citation flow between them.
For each year, we use Infomap with 100~optimization trials to search for the optimal multilevel solution.
We use the multilevel significance clustering approach described in the Methods section to assess the solution's robustness to slight perturbations in the data.
First, we create 1000 independent bootstrap networks by sampling each citation weight $w_{uv}$ from a Poisson distribution, $\hat{w}_{uv} \sim \text{Poisson}(w_{uv})$.
Then, we use Infomap to search for the optimal multilevel solution for each bootstrap network.
The bootstrap solutions have similar codelengths, with a variance of around $10^{-5}$.
Finally, we use the significance clustering algorithm to search for the largest fraction of nodes clustered together in at least a fraction~$p=0.95$ of the bootstrap solutions.

%

The resulting multilevel partitions organize science into research areas, further divided into research fields~(\cref{fig:science-alluvial}).
With the multilevel solution and unrecorded teleportation scheme, we do not exactly reproduce the results presented in Ref.~\onlinecite{rosvall2010mapping}.
The life sciences show higher diversification, with more significant research fields and lower citation flows in molecular- and cell biology containing J.~Biol.~Chem., Nature, PNAS, Science, Cell, and so on.

\subsection*{First- and second-order citation networks}

In the second case study, we visualize the effects of using higher-order network models with alluvial diagrams.
We organize the citation data from the Thomson-Reuters Journal Citation Reports into citation pathways.\cite{persson2016maps,wang2016large}
The data contain citations between articles published from 2007 to 2012 in the $10\,000$ journals with the highest impact factor, and all citation pathways contain at least one article published in 2009.
When aggregated to journals, we are left with $69,738,205$ weighted trigrams.

To study the effect of a second-order model, we model the data using both first- and second-order Markov chains.
We create a first-order network by discarding the first step from each trigram.
For example, the trigram $i\to j\to k$  with weight $w$ becomes the directed link $j\to k$ with the same weight,
resulting in 69 million links between the $10,000$ nodes.
Using the complete trigram data, we create a second-order network.
For each trigram $i\to j\to k$ with weight $w$, we create two state nodes if they do not already exist:
\begin{itemize}
    \item $\alpha_{i \to j}$ in physical node $j$ representing the memory of coming from $i$,
    \item $\alpha_{j \to k}$ in physical node $k$ representing the memory of coming from $j$.
\end{itemize}
We connect the state nodes with a directed link $\alpha_{i \to j} \to \alpha_{j \to k}$ with weight $w$.
The resulting second-order network has around $3.9$ million state nodes connected by 69 million links. 

Because the second-order network has two orders of magnitude more state nodes than the first-order network has physical nodes, the community detection search space is much larger, significantly impacting the computational time.
The first-order network takes around two minutes for ten optimization trials, while the second-order network takes around nine hours for the same task.
The resulting first-order partition has codelength $L^{(1)} = 8.44$~bits, five top modules, and four levels.
The second-order partition has codelength $L^{(2)} = 7.83$~bits, around $4,700$ top modules, and five levels.
Although the second-order partition has many top modules, most are tiny, containing only one or a few state nodes.
To downplay small modules at the fringe of the citation data, we compare the partition's effective number of top modules using the perplexity $M_{\text{eff}} = 2^{H(M)}$, with Shannon entropy
\begin{equation}
    H(M) = - \sum_m \pi_m \log_2 \pi_m,
\end{equation}
where $\pi_m = \sum_{i \in m} \pi_i$ is the total flow volume of the nodes $i$ in module $m$.
With this metric, the first- and second-order partitions are similar with $M_\text{eff}^{(1)} = 2.35$ and $M_\text{eff}^{(2)} = 2.73$ effective top modules, respectively.

After detecting communities, we aggregate redundant state nodes in the second-order network before visualization for better performance.
We lump state nodes in the same physical node and leaf module and aggregate their flows,
reducing the number of states to visualize from 3.9 million to 355 thousand.
After lumping, we remove any state nodes with zero flow that would not contribute to the alluvial diagram layout, further reducing the number of states to 271 thousand.
Then, we create pseudo-states in the first-order network to match the higher-order state nodes.
After this step, both networks contain 271~thousand state nodes. In the first-order network, all state nodes are in the same module as their physical node.

\begin{figure}[htp!]
    \centering
    \includegraphics[width=\linewidth]{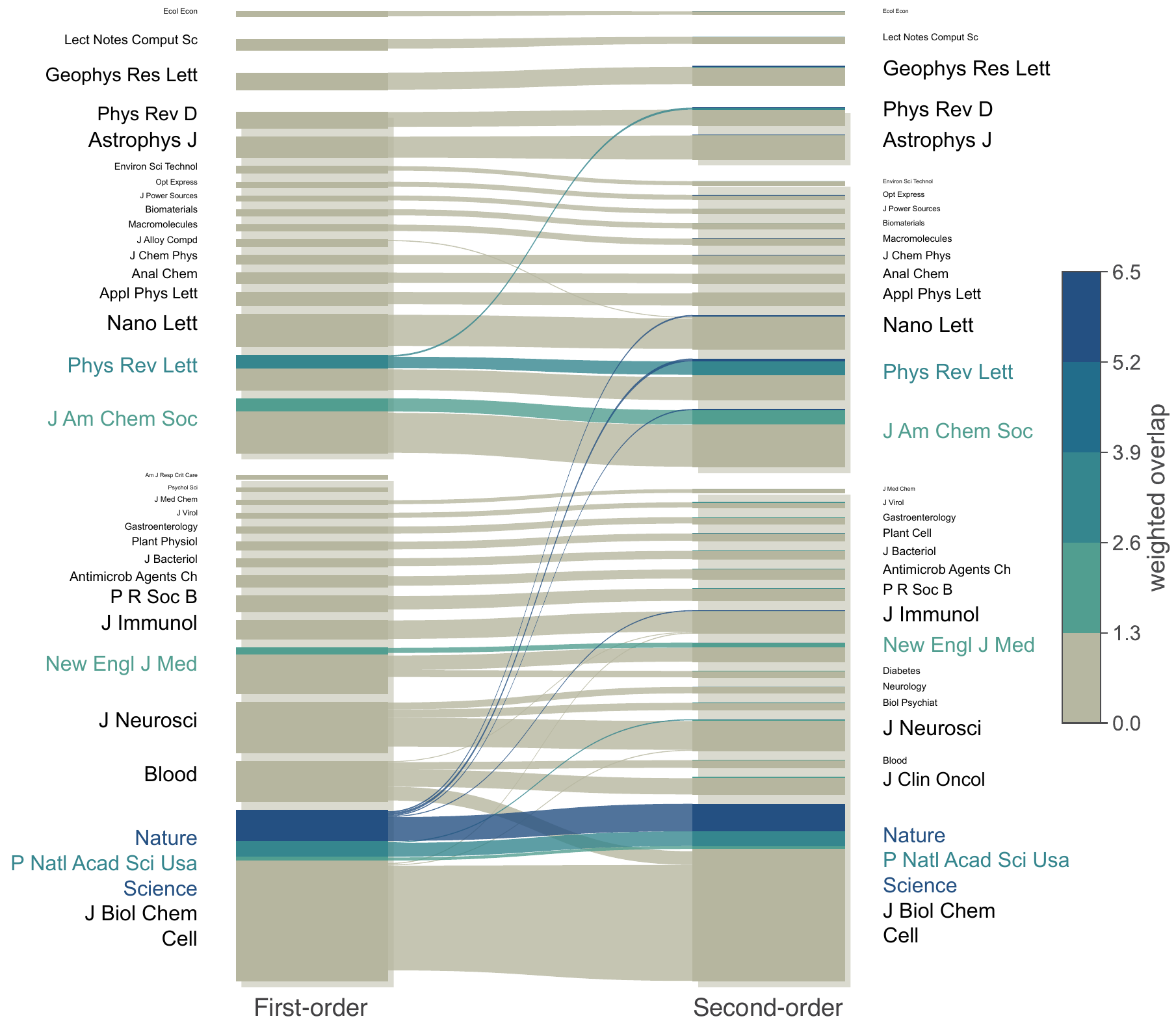}
    \caption{A second-order Markov model results in overlapping multidisciplinary journals.
    The leftmost network is first-order, and the rightmost is second-order.
    Colors indicate the journal's weighted overlap in the second-order network.}
    \label{fig:science-higher-order}
\end{figure}
The alluvial diagram shows how the second-order model separates cosmology and astrophysics from the physical sciences (\cref{fig:science-alluvial}).
The cell- and molecular biology submodule containing Nature, PNAS, and Science grows,
and the multidisciplinary journals' submodules in the life sciences divide into smaller modules.

\begin{figure}[H]
    \centering
    \includegraphics[width=\linewidth]{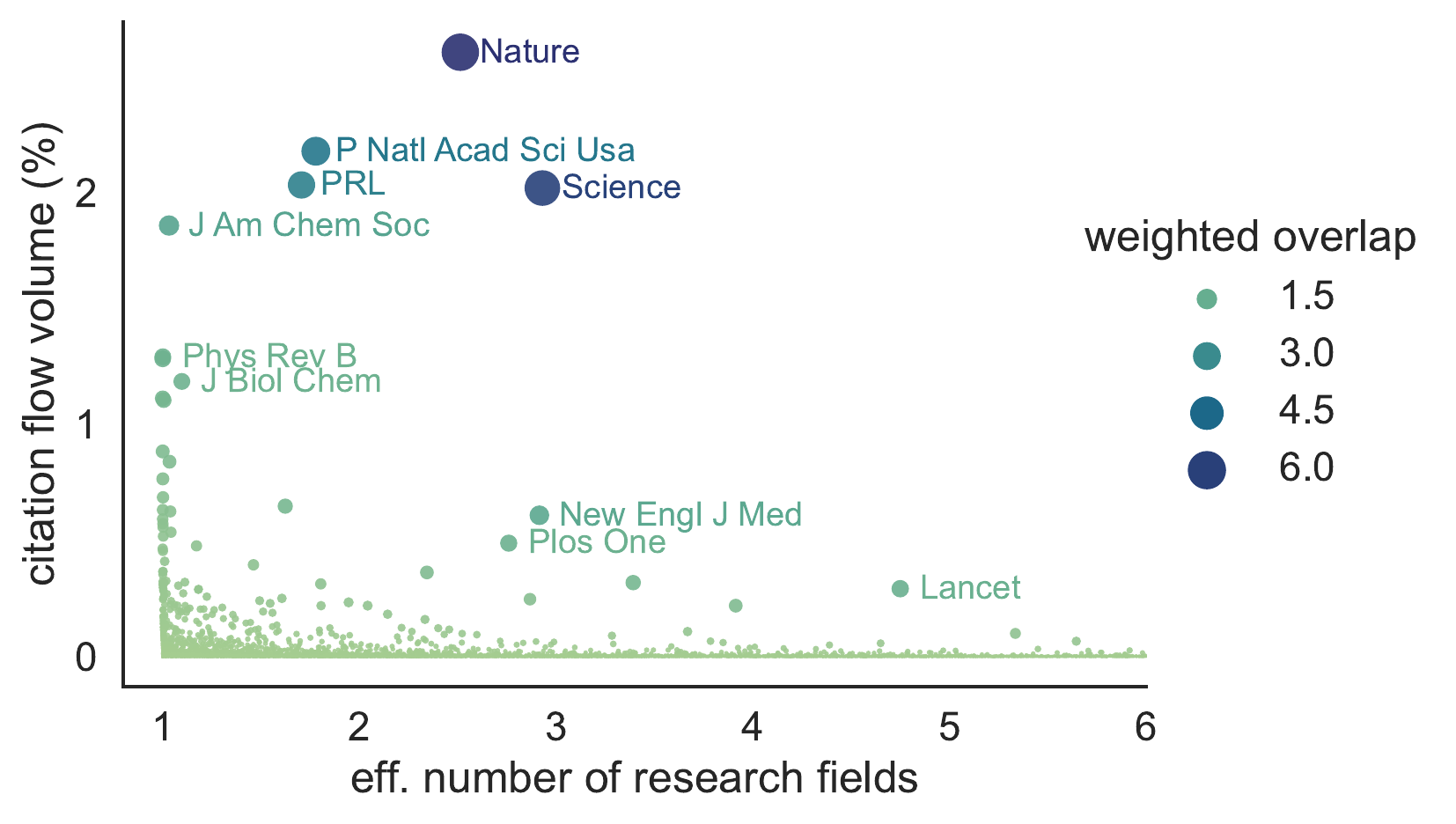}
    \caption{Influential multidisciplinary journals. The weighted overlap is the product of the journal's citation flow and its effective number of research fields. We limit the x-axis to six fields, but some small journals are in more than 30 research fields.}
    \label{fig:weighted-overlap}
\end{figure}
Above all, Nature, Science, and PNAS are all recognized as multidisciplinary journals represented in multiple research fields.
To quantify how a higher-order model captures their citation flows, we investigate in how many research fields journals are present.
Since a single research field dominates most journals' citation flows, we measure the effective number of research fields.
With journal $i$'s module-aggregated state node flow $\pmb{\pi}_i = \{ \pi_{i\alpha} \}$, we calculate its effective number of research fields $r_i = 2^{H(\pmb{\pi}_i)}$ with the entropy $H(\pmb{\pi}_i) = - \sum_\alpha \pi_{i\alpha} \log_2 \pi_{i\alpha}$.
%
%
With this metric, the most overlapping journals are Bratislava Medical J., Quality and Quantity, and Harvard Business Review -- tiny journals with only around $10^{-4}$ percent of the total citation flow.
To highlight prominent, multidisciplinary journals and mesoscale changes in the citation flows, we weigh each journal's effective number of research fields with its total citation flow $\pi_i = \sum_\alpha \pi_{i\alpha}$ for a weighted overlap
\begin{equation}
    o_i = r_i \pi_i.
\end{equation}
The journals with the highest weighted overlap are Nature, Science, and PNAS (\cref{fig:weighted-overlap}).

The life sciences contain more of the multidisciplinary citation flow than the other research areas.
By aggregating the weighted overlap $o_i$ on the leaf modules $m$,
\begin{equation}
    o_m = \sum_{i \in m} o_i,
\end{equation}
around 60 percent of the 1000 most overlapping leaf modules are in the life sciences, followed by the physical sciences with 18 percent.

\subsection*{Collaboration hypergraph using different representations}

\begin{figure*}[hpbt!]
    \centering
    \subfloat{\label{fig:schematic-hypergraph-a}}
    \subfloat{\label{fig:schematic-hypergraph-b}}
    \subfloat{\label{fig:schematic-hypergraph-c}}
    \vspace*{12pt}
    \begin{overpic}[width=0.9\textwidth]{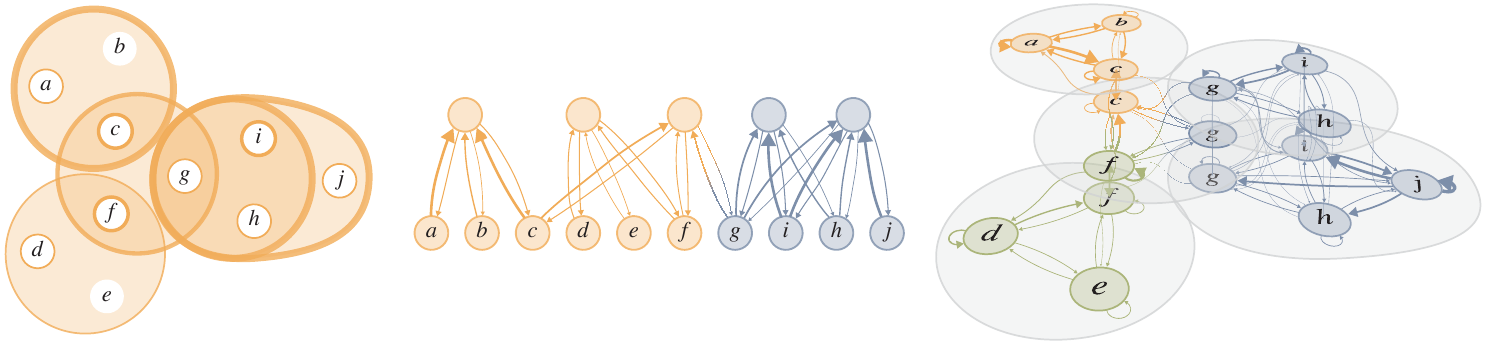}
        \put(0,23){\small\sffamily{\textbf{a}}}
        \put(29,23){\small\sffamily{\textbf{b}}}
        \put(64,23){\small\sffamily{\textbf{c}}}
    \end{overpic}
    \caption{Schematic hypergraph with edge-dependent node weights (a) and flow-equivalent network representations.
    b) A bipartite representation where hyperedges form hyperedge-nodes connecting all nodes in the hyperedge.
    c) A multilayer representation where each hyperedge forms a layer containing the hyperedge's nodes.
    The figure is adapted from
    Ref.~\onlinecite{eriksson2021choosing},
    licensed under \href{https://creativecommons.org/licenses/by/4.0/}{CC BY 4.0}.}
    \label{fig:schematic-hypergraph}
\end{figure*}

Finally, we study how a hypergraph's different first-order and multilayer network representations affect the detected communities.
We use a collaboration hypergraph extracted from the 734 references in the review article ``Networks beyond pairwise interactions: structure and dynamics.''\cite{eriksson2021choosing,battiston2020networks}
The referenced articles form hyperedges linking their authors.
These hyperedges overlap in those authors who authored multiple papers, with the largest connected component containing 361 author nodes in 220 hyperedges.

The hypergraph has article hyperedge weights $w(e) = \ln(c + 1) + 1$ where $c$ is the number of citations for that article in December 2020.\cite{eriksson2021choosing}
To model the author's unequal contributions to articles, we use hyperedge-dependent node weights\cite{chitra2019random}
\begin{equation}
    \gamma_e(i) = \begin{cases}
        2 \quad \text{if $i$ is first or last author},\\
        1 \quad \text{otherwise.}
    \end{cases}
\end{equation}
We weigh alphabetically sorted authors uniformly because their contributions are hard to determine.

From this hypergraph, we generate bipartite and multilayer hypergraph representations with identical node visit rates using the method described in Ref.\ \onlinecite{eriksson2021choosing} (\cref{fig:schematic-hypergraph}).
We also generate a multilayer network using a so-called hyperedge-similarity model that increases the probability of a random walk staying among similar hyperedges.\cite{eriksson2021choosing}
This model reinforces community structure with modules formed by similar sets of collaborators.
We let Infomap search for optimal multilevel solutions in the three network representations.
As before, we create pseudo-state nodes in the bipartite network to match them with the multilayer networks' state nodes.

\begin{figure*}[hpt!]
    \centering
    \includegraphics[width=0.85\textwidth]{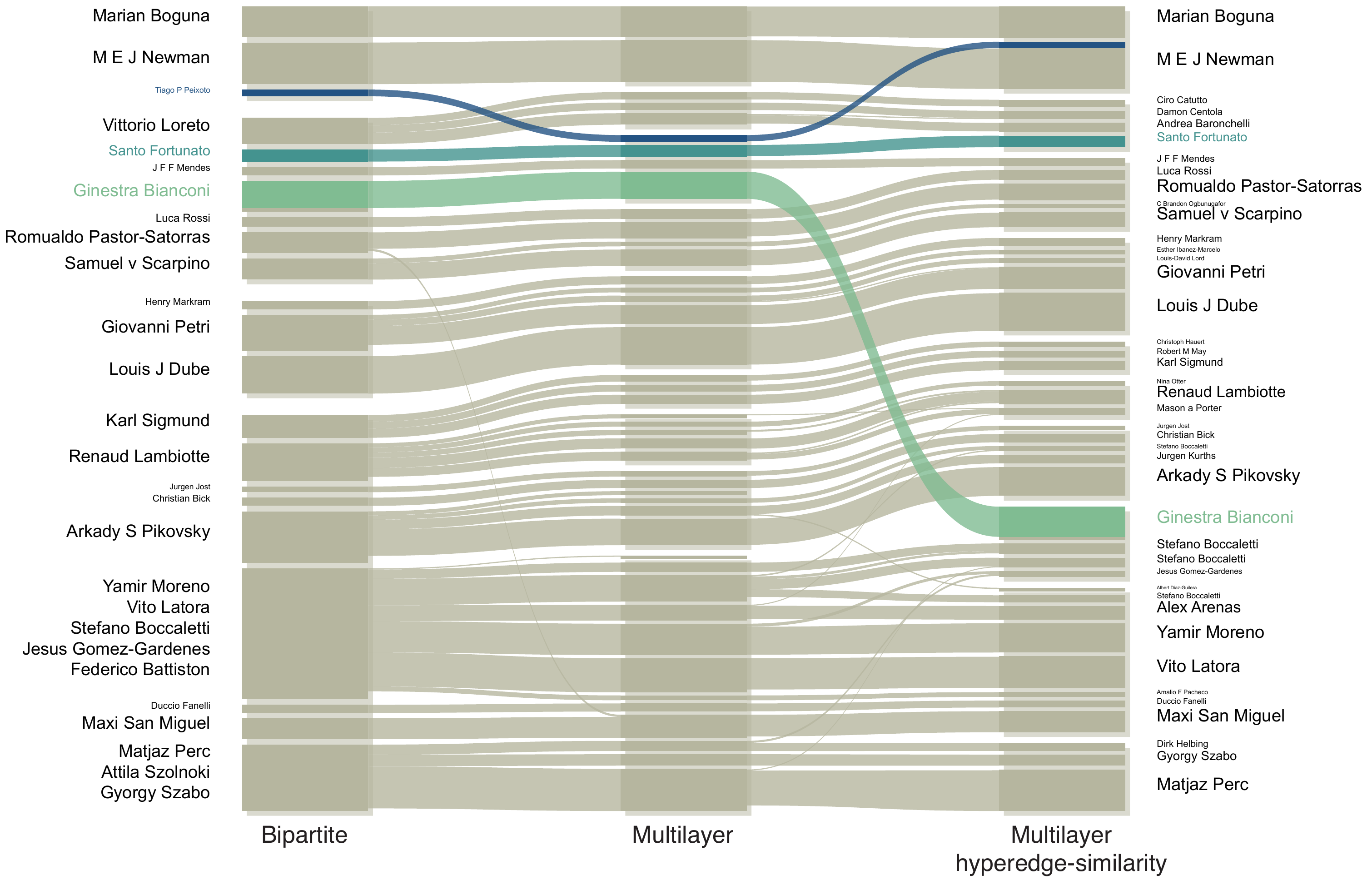}
    \caption{Network science collaboration hypergraph represented with two visit-rate-equivalent networks and a model that favors flow staying inside similar hyperedges. Modules are sorted to minimize overlap and show the names of their highest-ranking researchers. Here, we show the second-deepest level. Author groups that change top-level assignments in different networks are highlighted.}
    \label{fig:hypergraph-alluvial}
\end{figure*}

The resulting partitions have effectively three or four levels.
The top-level organization is most coarse-grained for the bipartite representation and most fine-grained for the hyperedge-similarity representation (\cref{fig:hypergraph-alluvial}). 
Only the multilayer representation assigns the submodule ``Peixoto'' together with the top module in which Bianconi is the highest-ranking author.
It also assigns Fortunato to a different top module than the hyperedge-similarity partition.
Finally, Bocaletti overlaps as the highest-ranking author in two submodules in the hyperedge-similarity partition in the same top module as Bianconi. 

\section*{Conclusions}

We have extended alluvial diagrams to higher-order networks with multilevel and overlapping communities and implemented an interactive web application available for anyone to use.
In three case studies, we have used alluvial diagrams to show how the multilevel organization of science changes over time, how a second-order model compares to a first-order model, and how different hypergraph-flow equivalent networks influence the flow of ideas among network scientists.

We have focused on flow-based community detection using the map equation framework and the search algorithm Infomap.
The generalized alluvial diagrams apply to any community-detection algorithm and are particularly relevant for simplifying and highlighting complex multilevel and overlapping modular descriptions of large higher-order networks.

\section*{Data and source code availability}

The interactive web application is available at {\small\url{https://www.mapequation.org/alluvial}}, and its source code at {\small\url{https://github.com/mapequation/alluvial-generator}}.
The multilevel significance clustering code is available at {\small\url{https://github.com/mapequation/multilevel-significance-clustering}}.
All other relevant data and code are available at {\small\url{https://github.com/mapequation/mapping-change-2}}.

\section*{Author contributions}

A.H.\ and M.R.\ devised the study.
A.H.\ performed the experiments.
A.H.\ and D.E.\ implemented the interactive web application.
A.H.\ and M.R.\ wrote the manuscript.
All authors edited and accepted the manuscript in its final form.

\section*{Competing interests}

The authors declare that they have no competing interests.

\acknowledgments

A.H.\ was supported by the Swedish Foundation for Strategic Research, Grant No. SB16-0089.
D.E.\ and M.R.\ were supported by the Swedish Research Council (2016-00796).

\bibliography{paper}

\begin{thebibliography}{40}
\expandafter\ifx\csname natexlab\endcsname\relax\def\natexlab#1{#1}\fi
\expandafter\ifx\csname bibnamefont\endcsname\relax
  \def\bibnamefont#1{#1}\fi
\expandafter\ifx\csname bibfnamefont\endcsname\relax
  \def\bibfnamefont#1{#1}\fi
\expandafter\ifx\csname citenamefont\endcsname\relax
  \def\citenamefont#1{#1}\fi
\expandafter\ifx\csname url\endcsname\relax
  \def\url#1{\texttt{#1}}\fi
\expandafter\ifx\csname urlprefix\endcsname\relax\def\urlprefix{URL }\fi
\providecommand{\bibinfo}[2]{#2}
\providecommand{\eprint}[2][]{\url{#2}}

\bibitem[{\citenamefont{Edler et~al.}(2017{\natexlab{a}})\citenamefont{Edler,
  Guedes, Zizka, Rosvall, and Antonelli}}]{edler2017infomap}
\bibinfo{author}{\bibfnamefont{D.}~\bibnamefont{Edler}},
  \bibinfo{author}{\bibfnamefont{T.}~\bibnamefont{Guedes}},
  \bibinfo{author}{\bibfnamefont{A.}~\bibnamefont{Zizka}},
  \bibinfo{author}{\bibfnamefont{M.}~\bibnamefont{Rosvall}}, \bibnamefont{and}
  \bibinfo{author}{\bibfnamefont{A.}~\bibnamefont{Antonelli}},
  \bibinfo{journal}{Systematic biology} \textbf{\bibinfo{volume}{66}},
  \bibinfo{pages}{197} (\bibinfo{year}{2017}{\natexlab{a}}).

\bibitem[{\citenamefont{Calatayud et~al.}(2020)\citenamefont{Calatayud,
  Andivia, Escudero, Meli{\'a}n, Bernardo-Madrid, Stoffel, Aponte, Medina,
  Molina-Venegas, Arnan et~al.}}]{calatayud2020positive}
\bibinfo{author}{\bibfnamefont{J.}~\bibnamefont{Calatayud}},
  \bibinfo{author}{\bibfnamefont{E.}~\bibnamefont{Andivia}},
  \bibinfo{author}{\bibfnamefont{A.}~\bibnamefont{Escudero}},
  \bibinfo{author}{\bibfnamefont{C.~J.} \bibnamefont{Meli{\'a}n}},
  \bibinfo{author}{\bibfnamefont{R.}~\bibnamefont{Bernardo-Madrid}},
  \bibinfo{author}{\bibfnamefont{M.}~\bibnamefont{Stoffel}},
  \bibinfo{author}{\bibfnamefont{C.}~\bibnamefont{Aponte}},
  \bibinfo{author}{\bibfnamefont{N.~G.} \bibnamefont{Medina}},
  \bibinfo{author}{\bibfnamefont{R.}~\bibnamefont{Molina-Venegas}},
  \bibinfo{author}{\bibfnamefont{X.}~\bibnamefont{Arnan}},
  \bibnamefont{et~al.}, \bibinfo{journal}{Nature ecology \& evolution}
  \textbf{\bibinfo{volume}{4}}, \bibinfo{pages}{40} (\bibinfo{year}{2020}).

\bibitem[{\citenamefont{Farage et~al.}(2021)\citenamefont{Farage, Edler,
  Ekl{\"o}f, Rosvall, and Pilosof}}]{farage2021identifying}
\bibinfo{author}{\bibfnamefont{C.}~\bibnamefont{Farage}},
  \bibinfo{author}{\bibfnamefont{D.}~\bibnamefont{Edler}},
  \bibinfo{author}{\bibfnamefont{A.}~\bibnamefont{Ekl{\"o}f}},
  \bibinfo{author}{\bibfnamefont{M.}~\bibnamefont{Rosvall}}, \bibnamefont{and}
  \bibinfo{author}{\bibfnamefont{S.}~\bibnamefont{Pilosof}},
  \bibinfo{journal}{Methods in Ecology and Evolution}
  \textbf{\bibinfo{volume}{12}}, \bibinfo{pages}{778} (\bibinfo{year}{2021}).

\bibitem[{\citenamefont{Calatayud et~al.}(2021)\citenamefont{Calatayud, Neuman,
  Rojas, Eriksson, and Rosvall}}]{joaquin2021regularities}
\bibinfo{author}{\bibfnamefont{J.}~\bibnamefont{Calatayud}},
  \bibinfo{author}{\bibfnamefont{M.}~\bibnamefont{Neuman}},
  \bibinfo{author}{\bibfnamefont{A.}~\bibnamefont{Rojas}},
  \bibinfo{author}{\bibfnamefont{A.}~\bibnamefont{Eriksson}}, \bibnamefont{and}
  \bibinfo{author}{\bibfnamefont{M.}~\bibnamefont{Rosvall}},
  \bibinfo{journal}{eLife} \textbf{\bibinfo{volume}{10}}
  (\bibinfo{year}{2021}).

\bibitem[{\citenamefont{Neuman}(2022)}]{neuman2022pisa}
\bibinfo{author}{\bibfnamefont{M.}~\bibnamefont{Neuman}},
  \bibinfo{journal}{Plos one} \textbf{\bibinfo{volume}{17}},
  \bibinfo{pages}{e0267040} (\bibinfo{year}{2022}).

\bibitem[{\citenamefont{Edler et~al.}(2022{\natexlab{a}})\citenamefont{Edler,
  Holmgren, Rojas, Rosvall, and Antonelli}}]{edler2022infomap}
\bibinfo{author}{\bibfnamefont{D.}~\bibnamefont{Edler}},
  \bibinfo{author}{\bibfnamefont{A.}~\bibnamefont{Holmgren}},
  \bibinfo{author}{\bibfnamefont{A.}~\bibnamefont{Rojas}},
  \bibinfo{author}{\bibfnamefont{M.}~\bibnamefont{Rosvall}}, \bibnamefont{and}
  \bibinfo{author}{\bibfnamefont{A.}~\bibnamefont{Antonelli}}
  (\bibinfo{year}{2022}{\natexlab{a}}).

\bibitem[{\citenamefont{Rojas et~al.}(2022)\citenamefont{Rojas, Eriksson,
  Neuman, Edler, Blocker, and Rosvall}}]{rojas2022natural}
\bibinfo{author}{\bibfnamefont{A.}~\bibnamefont{Rojas}},
  \bibinfo{author}{\bibfnamefont{A.}~\bibnamefont{Eriksson}},
  \bibinfo{author}{\bibfnamefont{M.}~\bibnamefont{Neuman}},
  \bibinfo{author}{\bibfnamefont{D.}~\bibnamefont{Edler}},
  \bibinfo{author}{\bibfnamefont{C.}~\bibnamefont{Blocker}}, \bibnamefont{and}
  \bibinfo{author}{\bibfnamefont{M.}~\bibnamefont{Rosvall}},
  \bibinfo{journal}{bioRxiv}  (\bibinfo{year}{2022}).

\bibitem[{\citenamefont{Rosvall and Bergstrom}(2008)}]{rosvall2008maps}
\bibinfo{author}{\bibfnamefont{M.}~\bibnamefont{Rosvall}} \bibnamefont{and}
  \bibinfo{author}{\bibfnamefont{C.~T.} \bibnamefont{Bergstrom}},
  \bibinfo{journal}{Proceedings of the National Academy of Sciences}
  \textbf{\bibinfo{volume}{105}}, \bibinfo{pages}{1118} (\bibinfo{year}{2008}).

\bibitem[{\citenamefont{Fortunato}(2010)}]{fortunato2010community}
\bibinfo{author}{\bibfnamefont{S.}~\bibnamefont{Fortunato}},
  \bibinfo{journal}{Physics reports} \textbf{\bibinfo{volume}{486}},
  \bibinfo{pages}{75} (\bibinfo{year}{2010}).

\bibitem[{\citenamefont{Schaub et~al.}(2017)\citenamefont{Schaub, Delvenne,
  Rosvall, and Lambiotte}}]{schaub2017many}
\bibinfo{author}{\bibfnamefont{M.~T.} \bibnamefont{Schaub}},
  \bibinfo{author}{\bibfnamefont{J.-C.} \bibnamefont{Delvenne}},
  \bibinfo{author}{\bibfnamefont{M.}~\bibnamefont{Rosvall}}, \bibnamefont{and}
  \bibinfo{author}{\bibfnamefont{R.}~\bibnamefont{Lambiotte}},
  \bibinfo{journal}{Applied network science} \textbf{\bibinfo{volume}{2}},
  \bibinfo{pages}{1} (\bibinfo{year}{2017}).

\bibitem[{\citenamefont{Traag et~al.}(2019)\citenamefont{Traag, Waltman, and
  Van~Eck}}]{traag2019louvain}
\bibinfo{author}{\bibfnamefont{V.~A.} \bibnamefont{Traag}},
  \bibinfo{author}{\bibfnamefont{L.}~\bibnamefont{Waltman}}, \bibnamefont{and}
  \bibinfo{author}{\bibfnamefont{N.~J.} \bibnamefont{Van~Eck}},
  \bibinfo{journal}{Scientific reports} \textbf{\bibinfo{volume}{9}},
  \bibinfo{pages}{1} (\bibinfo{year}{2019}).

\bibitem[{\citenamefont{Peixoto}(2019)}]{peixoto2019bayesian}
\bibinfo{author}{\bibfnamefont{T.~P.} \bibnamefont{Peixoto}},
  \bibinfo{journal}{Advances in network clustering and blockmodeling} pp.
  \bibinfo{pages}{289--332} (\bibinfo{year}{2019}).

\bibitem[{\citenamefont{Danon et~al.}(2005)\citenamefont{Danon, Diaz-Guilera,
  Duch, and Arenas}}]{danon2005comparing}
\bibinfo{author}{\bibfnamefont{L.}~\bibnamefont{Danon}},
  \bibinfo{author}{\bibfnamefont{A.}~\bibnamefont{Diaz-Guilera}},
  \bibinfo{author}{\bibfnamefont{J.}~\bibnamefont{Duch}}, \bibnamefont{and}
  \bibinfo{author}{\bibfnamefont{A.}~\bibnamefont{Arenas}},
  \bibinfo{journal}{Journal of statistical mechanics: Theory and experiment}
  \textbf{\bibinfo{volume}{2005}}, \bibinfo{pages}{P09008}
  (\bibinfo{year}{2005}).

\bibitem[{\citenamefont{Amelio and Pizzuti}(2017)}]{amelio2017correction}
\bibinfo{author}{\bibfnamefont{A.}~\bibnamefont{Amelio}} \bibnamefont{and}
  \bibinfo{author}{\bibfnamefont{C.}~\bibnamefont{Pizzuti}},
  \bibinfo{journal}{Computational Intelligence} \textbf{\bibinfo{volume}{33}},
  \bibinfo{pages}{579} (\bibinfo{year}{2017}).

\bibitem[{\citenamefont{Newman et~al.}(2020)\citenamefont{Newman, Cantwell, and
  Young}}]{newman2020improved}
\bibinfo{author}{\bibfnamefont{M.~E.} \bibnamefont{Newman}},
  \bibinfo{author}{\bibfnamefont{G.~T.} \bibnamefont{Cantwell}},
  \bibnamefont{and} \bibinfo{author}{\bibfnamefont{J.-G.} \bibnamefont{Young}},
  \bibinfo{journal}{Physical Review E} \textbf{\bibinfo{volume}{101}},
  \bibinfo{pages}{042304} (\bibinfo{year}{2020}).

\bibitem[{\citenamefont{Rosvall and Bergstrom}(2010)}]{rosvall2010mapping}
\bibinfo{author}{\bibfnamefont{M.}~\bibnamefont{Rosvall}} \bibnamefont{and}
  \bibinfo{author}{\bibfnamefont{C.~T.} \bibnamefont{Bergstrom}},
  \bibinfo{journal}{PloS one} \textbf{\bibinfo{volume}{5}},
  \bibinfo{pages}{e8694} (\bibinfo{year}{2010}).

\bibitem[{\citenamefont{Liu et~al.}(2013)\citenamefont{Liu, Derudder,
  Csom{\'o}s, and Taylor}}]{liu2013featured}
\bibinfo{author}{\bibfnamefont{X.}~\bibnamefont{Liu}},
  \bibinfo{author}{\bibfnamefont{B.}~\bibnamefont{Derudder}},
  \bibinfo{author}{\bibfnamefont{G.}~\bibnamefont{Csom{\'o}s}},
  \bibnamefont{and} \bibinfo{author}{\bibfnamefont{P.}~\bibnamefont{Taylor}},
  \bibinfo{journal}{Environment and Planning A} \textbf{\bibinfo{volume}{45}},
  \bibinfo{pages}{1005} (\bibinfo{year}{2013}).

\bibitem[{\citenamefont{Ruan et~al.}(2017)\citenamefont{Ruan, Hou, and
  Hu}}]{ruan2017detecting}
\bibinfo{author}{\bibfnamefont{W.}~\bibnamefont{Ruan}},
  \bibinfo{author}{\bibfnamefont{H.}~\bibnamefont{Hou}}, \bibnamefont{and}
  \bibinfo{author}{\bibfnamefont{Z.}~\bibnamefont{Hu}},
  \bibinfo{journal}{Journal of Data and Information Science}
  \textbf{\bibinfo{volume}{2}}, \bibinfo{pages}{37} (\bibinfo{year}{2017}).

\bibitem[{\citenamefont{Pal et~al.}(2022)\citenamefont{Pal, Chopra, Awasthi,
  Bandhey, Nagori, Sethi et~al.}}]{pal2022predicting}
\bibinfo{author}{\bibfnamefont{R.}~\bibnamefont{Pal}},
  \bibinfo{author}{\bibfnamefont{H.}~\bibnamefont{Chopra}},
  \bibinfo{author}{\bibfnamefont{R.}~\bibnamefont{Awasthi}},
  \bibinfo{author}{\bibfnamefont{H.}~\bibnamefont{Bandhey}},
  \bibinfo{author}{\bibfnamefont{A.}~\bibnamefont{Nagori}},
  \bibinfo{author}{\bibfnamefont{T.}~\bibnamefont{Sethi}},
  \bibnamefont{et~al.}, \bibinfo{journal}{Journal of medical Internet research}
  \textbf{\bibinfo{volume}{24}}, \bibinfo{pages}{e34067}
  (\bibinfo{year}{2022}).

\bibitem[{\citenamefont{Remy et~al.}(2017)\citenamefont{Remy, Rym, and
  Matthieu}}]{remy2017tracking}
\bibinfo{author}{\bibfnamefont{C.}~\bibnamefont{Remy}},
  \bibinfo{author}{\bibfnamefont{B.}~\bibnamefont{Rym}}, \bibnamefont{and}
  \bibinfo{author}{\bibfnamefont{L.}~\bibnamefont{Matthieu}}, in
  \emph{\bibinfo{booktitle}{International conference on complex networks and
  their applications}} (\bibinfo{organization}{Springer},
  \bibinfo{year}{2017}), pp. \bibinfo{pages}{166--177}.

\bibitem[{\citenamefont{Petrun~Sayers et~al.}(2021)\citenamefont{Petrun~Sayers,
  Parker, Seelam, and Finucane}}]{petrun2021disasters}
\bibinfo{author}{\bibfnamefont{E.~L.} \bibnamefont{Petrun~Sayers}},
  \bibinfo{author}{\bibfnamefont{A.~M.} \bibnamefont{Parker}},
  \bibinfo{author}{\bibfnamefont{R.}~\bibnamefont{Seelam}}, \bibnamefont{and}
  \bibinfo{author}{\bibfnamefont{M.~L.} \bibnamefont{Finucane}},
  \bibinfo{journal}{Journal of Contingencies and Crisis Management}
  \textbf{\bibinfo{volume}{29}}, \bibinfo{pages}{342} (\bibinfo{year}{2021}).

\bibitem[{\citenamefont{Kivel{\"a} et~al.}(2014)\citenamefont{Kivel{\"a},
  Arenas, Barthelemy, Gleeson, Moreno, and Porter}}]{kivela2014multilayer}
\bibinfo{author}{\bibfnamefont{M.}~\bibnamefont{Kivel{\"a}}},
  \bibinfo{author}{\bibfnamefont{A.}~\bibnamefont{Arenas}},
  \bibinfo{author}{\bibfnamefont{M.}~\bibnamefont{Barthelemy}},
  \bibinfo{author}{\bibfnamefont{J.~P.} \bibnamefont{Gleeson}},
  \bibinfo{author}{\bibfnamefont{Y.}~\bibnamefont{Moreno}}, \bibnamefont{and}
  \bibinfo{author}{\bibfnamefont{M.~A.} \bibnamefont{Porter}},
  \bibinfo{journal}{Journal of complex networks} \textbf{\bibinfo{volume}{2}},
  \bibinfo{pages}{203} (\bibinfo{year}{2014}).

\bibitem[{\citenamefont{Rosvall et~al.}(2014)\citenamefont{Rosvall, Esquivel,
  Lancichinetti, West, and Lambiotte}}]{rosvall2014memory}
\bibinfo{author}{\bibfnamefont{M.}~\bibnamefont{Rosvall}},
  \bibinfo{author}{\bibfnamefont{A.~V.} \bibnamefont{Esquivel}},
  \bibinfo{author}{\bibfnamefont{A.}~\bibnamefont{Lancichinetti}},
  \bibinfo{author}{\bibfnamefont{J.~D.} \bibnamefont{West}}, \bibnamefont{and}
  \bibinfo{author}{\bibfnamefont{R.}~\bibnamefont{Lambiotte}},
  \bibinfo{journal}{Nature communications} \textbf{\bibinfo{volume}{5}},
  \bibinfo{pages}{1} (\bibinfo{year}{2014}).

\bibitem[{\citenamefont{De~Domenico et~al.}(2015)\citenamefont{De~Domenico,
  Lancichinetti, Arenas, and Rosvall}}]{de2015identifying}
\bibinfo{author}{\bibfnamefont{M.}~\bibnamefont{De~Domenico}},
  \bibinfo{author}{\bibfnamefont{A.}~\bibnamefont{Lancichinetti}},
  \bibinfo{author}{\bibfnamefont{A.}~\bibnamefont{Arenas}}, \bibnamefont{and}
  \bibinfo{author}{\bibfnamefont{M.}~\bibnamefont{Rosvall}},
  \bibinfo{journal}{Physical Review X} \textbf{\bibinfo{volume}{5}},
  \bibinfo{pages}{011027} (\bibinfo{year}{2015}).

\bibitem[{\citenamefont{De~Domenico et~al.}(2016)\citenamefont{De~Domenico,
  Granell, Porter, and Arenas}}]{de2016physics}
\bibinfo{author}{\bibfnamefont{M.}~\bibnamefont{De~Domenico}},
  \bibinfo{author}{\bibfnamefont{C.}~\bibnamefont{Granell}},
  \bibinfo{author}{\bibfnamefont{M.~A.} \bibnamefont{Porter}},
  \bibnamefont{and} \bibinfo{author}{\bibfnamefont{A.}~\bibnamefont{Arenas}},
  \bibinfo{journal}{Nature Physics} \textbf{\bibinfo{volume}{12}},
  \bibinfo{pages}{901} (\bibinfo{year}{2016}).

\bibitem[{\citenamefont{Xu et~al.}(2016)\citenamefont{Xu, Wickramarathne, and
  Chawla}}]{xu2016representing}
\bibinfo{author}{\bibfnamefont{J.}~\bibnamefont{Xu}},
  \bibinfo{author}{\bibfnamefont{T.~L.} \bibnamefont{Wickramarathne}},
  \bibnamefont{and} \bibinfo{author}{\bibfnamefont{N.~V.}
  \bibnamefont{Chawla}}, \bibinfo{journal}{Science advances}
  \textbf{\bibinfo{volume}{2}}, \bibinfo{pages}{e1600028}
  (\bibinfo{year}{2016}).

\bibitem[{\citenamefont{Lambiotte et~al.}(2019)\citenamefont{Lambiotte,
  Rosvall, and Scholtes}}]{lambiotte2019networks}
\bibinfo{author}{\bibfnamefont{R.}~\bibnamefont{Lambiotte}},
  \bibinfo{author}{\bibfnamefont{M.}~\bibnamefont{Rosvall}}, \bibnamefont{and}
  \bibinfo{author}{\bibfnamefont{I.}~\bibnamefont{Scholtes}},
  \bibinfo{journal}{Nature physics} \textbf{\bibinfo{volume}{15}},
  \bibinfo{pages}{313} (\bibinfo{year}{2019}).

\bibitem[{\citenamefont{Rosvall and Bergstrom}(2011)}]{rosvall2011multilevel}
\bibinfo{author}{\bibfnamefont{M.}~\bibnamefont{Rosvall}} \bibnamefont{and}
  \bibinfo{author}{\bibfnamefont{C.~T.} \bibnamefont{Bergstrom}},
  \bibinfo{journal}{PloS one} \textbf{\bibinfo{volume}{6}},
  \bibinfo{pages}{e18209} (\bibinfo{year}{2011}).

\bibitem[{\citenamefont{Peixoto}(2014)}]{peixoto2014hierarchical}
\bibinfo{author}{\bibfnamefont{T.~P.} \bibnamefont{Peixoto}},
  \bibinfo{journal}{Physical Review X} \textbf{\bibinfo{volume}{4}},
  \bibinfo{pages}{011047} (\bibinfo{year}{2014}).

\bibitem[{\citenamefont{Holmgren
  et~al.}(2022{\natexlab{a}})\citenamefont{Holmgren, Edler, and
  Rosvall}}]{mapequation2022alluvial}
\bibinfo{author}{\bibfnamefont{A.}~\bibnamefont{Holmgren}},
  \bibinfo{author}{\bibfnamefont{D.}~\bibnamefont{Edler}}, \bibnamefont{and}
  \bibinfo{author}{\bibfnamefont{M.}~\bibnamefont{Rosvall}},
  \emph{\bibinfo{title}{{The MapEquation Alluvial Diagram Generator}}}
  (\bibinfo{year}{2022}{\natexlab{a}}),
  \urlprefix\url{https://mapequation.org/alluvial}.

\bibitem[{\citenamefont{Edler et~al.}(2022{\natexlab{b}})\citenamefont{Edler,
  Holmgren, and Rosvall}}]{mapequation2022software}
\bibinfo{author}{\bibfnamefont{D.}~\bibnamefont{Edler}},
  \bibinfo{author}{\bibfnamefont{A.}~\bibnamefont{Holmgren}}, \bibnamefont{and}
  \bibinfo{author}{\bibfnamefont{M.}~\bibnamefont{Rosvall}},
  \emph{\bibinfo{title}{{The MapEquation software package}}}
  (\bibinfo{year}{2022}{\natexlab{b}}),
  \urlprefix\url{https://mapequation.org}.

\bibitem[{\citenamefont{Edler et~al.}(2017{\natexlab{b}})\citenamefont{Edler,
  Bohlin et~al.}}]{edler2017mapping}
\bibinfo{author}{\bibfnamefont{D.}~\bibnamefont{Edler}},
  \bibinfo{author}{\bibfnamefont{L.}~\bibnamefont{Bohlin}},
  \bibnamefont{et~al.}, \bibinfo{journal}{Algorithms}
  \textbf{\bibinfo{volume}{10}}, \bibinfo{pages}{112}
  (\bibinfo{year}{2017}{\natexlab{b}}).

\bibitem[{\citenamefont{Zakai}(2011)}]{zakai2011emscripten}
\bibinfo{author}{\bibfnamefont{A.}~\bibnamefont{Zakai}}, in
  \emph{\bibinfo{booktitle}{Proceedings of the ACM international conference
  companion on Object oriented programming systems languages and applications
  companion}} (\bibinfo{year}{2011}), pp. \bibinfo{pages}{301--312}.

\bibitem[{\citenamefont{Holmgren
  et~al.}(2022{\natexlab{b}})\citenamefont{Holmgren, Edler, and
  Rosvall}}]{mapequation2022infomaponline}
\bibinfo{author}{\bibfnamefont{A.}~\bibnamefont{Holmgren}},
  \bibinfo{author}{\bibfnamefont{D.}~\bibnamefont{Edler}}, \bibnamefont{and}
  \bibinfo{author}{\bibfnamefont{M.}~\bibnamefont{Rosvall}},
  \emph{\bibinfo{title}{{Infomap Online}}}
  (\bibinfo{year}{2022}{\natexlab{b}}),
  \urlprefix\url{https://mapequation.org/infomap}.

\bibitem[{\citenamefont{Lambiotte and Rosvall}(2012)}]{lambiotte2012ranking}
\bibinfo{author}{\bibfnamefont{R.}~\bibnamefont{Lambiotte}} \bibnamefont{and}
  \bibinfo{author}{\bibfnamefont{M.}~\bibnamefont{Rosvall}},
  \bibinfo{journal}{Phys. Rev. E} \textbf{\bibinfo{volume}{85}},
  \bibinfo{pages}{056107} (\bibinfo{year}{2012}).

\bibitem[{\citenamefont{Persson et~al.}(2016)\citenamefont{Persson, Bohlin,
  Edler, and Rosvall}}]{persson2016maps}
\bibinfo{author}{\bibfnamefont{C.}~\bibnamefont{Persson}},
  \bibinfo{author}{\bibfnamefont{L.}~\bibnamefont{Bohlin}},
  \bibinfo{author}{\bibfnamefont{D.}~\bibnamefont{Edler}}, \bibnamefont{and}
  \bibinfo{author}{\bibfnamefont{M.}~\bibnamefont{Rosvall}},
  \bibinfo{journal}{arXiv preprint arXiv:1606.08328}  (\bibinfo{year}{2016}).

\bibitem[{\citenamefont{Wang and Waltman}(2016)}]{wang2016large}
\bibinfo{author}{\bibfnamefont{Q.}~\bibnamefont{Wang}} \bibnamefont{and}
  \bibinfo{author}{\bibfnamefont{L.}~\bibnamefont{Waltman}},
  \bibinfo{journal}{Journal of informetrics} \textbf{\bibinfo{volume}{10}},
  \bibinfo{pages}{347} (\bibinfo{year}{2016}).

\bibitem[{\citenamefont{Eriksson et~al.}(2021)\citenamefont{Eriksson, Edler,
  Rojas, de~Domenico, and Rosvall}}]{eriksson2021choosing}
\bibinfo{author}{\bibfnamefont{A.}~\bibnamefont{Eriksson}},
  \bibinfo{author}{\bibfnamefont{D.}~\bibnamefont{Edler}},
  \bibinfo{author}{\bibfnamefont{A.}~\bibnamefont{Rojas}},
  \bibinfo{author}{\bibfnamefont{M.}~\bibnamefont{de~Domenico}},
  \bibnamefont{and} \bibinfo{author}{\bibfnamefont{M.}~\bibnamefont{Rosvall}},
  \bibinfo{journal}{Communications Physics} \textbf{\bibinfo{volume}{4}},
  \bibinfo{pages}{1} (\bibinfo{year}{2021}).

\bibitem[{\citenamefont{Battiston et~al.}(2020)\citenamefont{Battiston,
  Cencetti, Iacopini, Latora, Lucas, Patania, Young, and
  Petri}}]{battiston2020networks}
\bibinfo{author}{\bibfnamefont{F.}~\bibnamefont{Battiston}},
  \bibinfo{author}{\bibfnamefont{G.}~\bibnamefont{Cencetti}},
  \bibinfo{author}{\bibfnamefont{I.}~\bibnamefont{Iacopini}},
  \bibinfo{author}{\bibfnamefont{V.}~\bibnamefont{Latora}},
  \bibinfo{author}{\bibfnamefont{M.}~\bibnamefont{Lucas}},
  \bibinfo{author}{\bibfnamefont{A.}~\bibnamefont{Patania}},
  \bibinfo{author}{\bibfnamefont{J.-G.} \bibnamefont{Young}}, \bibnamefont{and}
  \bibinfo{author}{\bibfnamefont{G.}~\bibnamefont{Petri}},
  \bibinfo{journal}{Physics Reports} \textbf{\bibinfo{volume}{874}},
  \bibinfo{pages}{1} (\bibinfo{year}{2020}).

\bibitem[{\citenamefont{Chitra and Raphael}(2019)}]{chitra2019random}
\bibinfo{author}{\bibfnamefont{U.}~\bibnamefont{Chitra}} \bibnamefont{and}
  \bibinfo{author}{\bibfnamefont{B.}~\bibnamefont{Raphael}}, in
  \emph{\bibinfo{booktitle}{International Conference on Machine Learning}}
  (\bibinfo{organization}{PMLR}, \bibinfo{year}{2019}), pp.
  \bibinfo{pages}{1172--1181}.

\end{thebibliography}

\balancecolsandclearpage

\renewcommand{\thesection}{SI}
\renewcommand{\thesubsection}{\arabic{subsection}}
\setcounter{figure}{0}
\renewcommand{\thefigure}{S\arabic{figure}}
\setcounter{lstlisting}{0}
\renewcommand{\thelstlisting}{S\arabic{lstlisting}}

\section{Supplementary information}

\subsection{Input formats}
\label{sec:input}

The alluvial diagram generator accepts the following input formats: \texttt{clu}, \texttt{tree}, and \texttt{json} (\cref{lst:clu,lst:clu_states,lst:clu_multilayer,lst:tree,lst:tree_states,lst:tree_multilayer,lst:json,lst:json_states,lst:json_multilayer}). It also accepts the \texttt{ftree} format, which is the same as the \texttt{tree} format with aggregated intra-module links, which are not used by the web application. In addition, it also understands any network format readable by Infomap and any combinations of the above files compressed as a \texttt{zip} file.

The ``flow'' column in the \texttt{clu} format can be substituted with any node weighting scheme to make the web application available to more researchers.

\begin{lstlisting}[language=Python,label={lst:clu},caption=\texttt{clu} format]
# node_id module flow
1 1 0.214286
2 1 0.142857
3 1 0.142857
4 2 0.214286
5 2 0.142857
6 2 0.142857
\end{lstlisting}

\begin{lstlisting}[language=Python,label={lst:clu_states},caption=\texttt{clu} states format]
# state_id module flow node_id
1 1 0.166667 1
2 1 0.166667 2
3 1 0.166667 3
4 2 0.166667 1
5 2 0.166667 4
6 2 0.166667 5
\end{lstlisting}

\begin{lstlisting}[language=Python,label={lst:clu_multilayer},caption=\texttt{clu} multilayer format]
# state_id module flow node_id layer_id
3 1 0.166667 1 2
4 1 0.166667 2 2
5 1 0.166667 3 2
0 2 0.166667 1 1
1 2 0.166667 4 1
2 2 0.166667 5 1
\end{lstlisting}

\begin{lstlisting}[language=Python,label={lst:tree},caption=\texttt{tree} format]
# path flow name node_id
1:1 0.214286 "1" 1
1:2 0.142857 "2" 2
1:3 0.142857 "3" 3
2:1 0.214286 "4" 4
2:2 0.142857 "5" 5
2:3 0.142857 "6" 6
\end{lstlisting}

\begin{lstlisting}[language=Python,label={lst:tree_states},caption=\texttt{tree} states format]
# path flow name state_id node_id
1:1 0.166667 "i" 1 1
1:2 0.166667 "j" 2 2
1:3 0.166667 "k" 3 3
2:1 0.166667 "i" 4 1
2:2 0.166667 "l" 5 4
2:3 0.166667 "m" 6 5
\end{lstlisting}

\begin{lstlisting}[language=Python,label={lst:tree_multilayer},caption=\texttt{tree} multilayer format]
# path flow name state_id node_id layer_id
1:1 0.166667 "i" 3 1 2
1:2 0.166667 "j" 4 2 2
1:3 0.166667 "k" 5 3 2
2:1 0.166667 "i" 0 1 1
2:2 0.166667 "l" 1 4 1
2:3 0.166667 "m" 2 5 1
\end{lstlisting}

\begin{lstlisting}[language=C,label={lst:json},caption=\texttt{json} format]
{
  "nodes": [
    { "path": [1, 1], "name": "1", "flow": 0.214286, "id": 1 },
    { "path": [1, 2], "name": "2", "flow": 0.142857, "id": 2 },
    { "path": [1, 3], "name": "3", "flow": 0.142857, "id": 3 },
    { "path": [2, 1], "name": "4", "flow": 0.214286, "id": 4 },
    { "path": [2, 2], "name": "5", "flow": 0.142857, "id": 5 },
    { "path": [2, 3], "name": "6", "flow": 0.142857, "id": 6 }
  ]
}
\end{lstlisting}

\begin{lstlisting}[language=C,label={lst:json_states},caption=\texttt{json} states format]
{
  "nodes": [
    { "path": [1, 1], "name": "i", "flow": 0.166667, "stateId": 1, "id": 1 },
    { "path": [1, 2], "name": "j", "flow": 0.166667, "stateId": 2, "id": 2 },
    { "path": [1, 3], "name": "k", "flow": 0.166667, "stateId": 3, "id": 3 },
    { "path": [2, 1], "name": "i", "flow": 0.166667, "stateId": 4, "id": 1 },
    { "path": [2, 2], "name": "l", "flow": 0.166667, "stateId": 5, "id": 4 },
    { "path": [2, 3], "name": "m", "flow": 0.166667, "stateId": 6, "id": 5 }
  ]
}
\end{lstlisting}

\begin{lstlisting}[language=C,label={lst:json_multilayer},caption=\texttt{json} multilayer format]
{
  "nodes": [
    { "path": [1, 1], "name": "i", "flow": 0.166667,
      "stateId": 3, "layerId": 2, "id": 1 },
    { "path": [1, 2], "name": "j", "flow": 0.166667,
      "stateId": 4, "layerId": 2, "id": 2 },
    { "path": [1, 3], "name": "k", "flow": 0.166667,
      "stateId": 5, "layerId": 2, "id": 3 },
    { "path": [2, 1], "name": "i", "flow": 0.166667,
      "stateId": 0, "layerId": 1, "id": 1 },
    { "path": [2, 2], "name": "l", "flow": 0.166667,
      "stateId": 1, "layerId": 1, "id": 4 },
    { "path": [2, 3], "name": "m", "flow": 0.166667,
      "stateId": 2, "layerId": 1, "id": 5 }
  ]
}
\end{lstlisting}

\onecolumngrid
\clearpage
\subsection{Supplementary figures}

\begin{figure*}[htp!]
    \centering
    \includegraphics[width=0.95\textwidth]{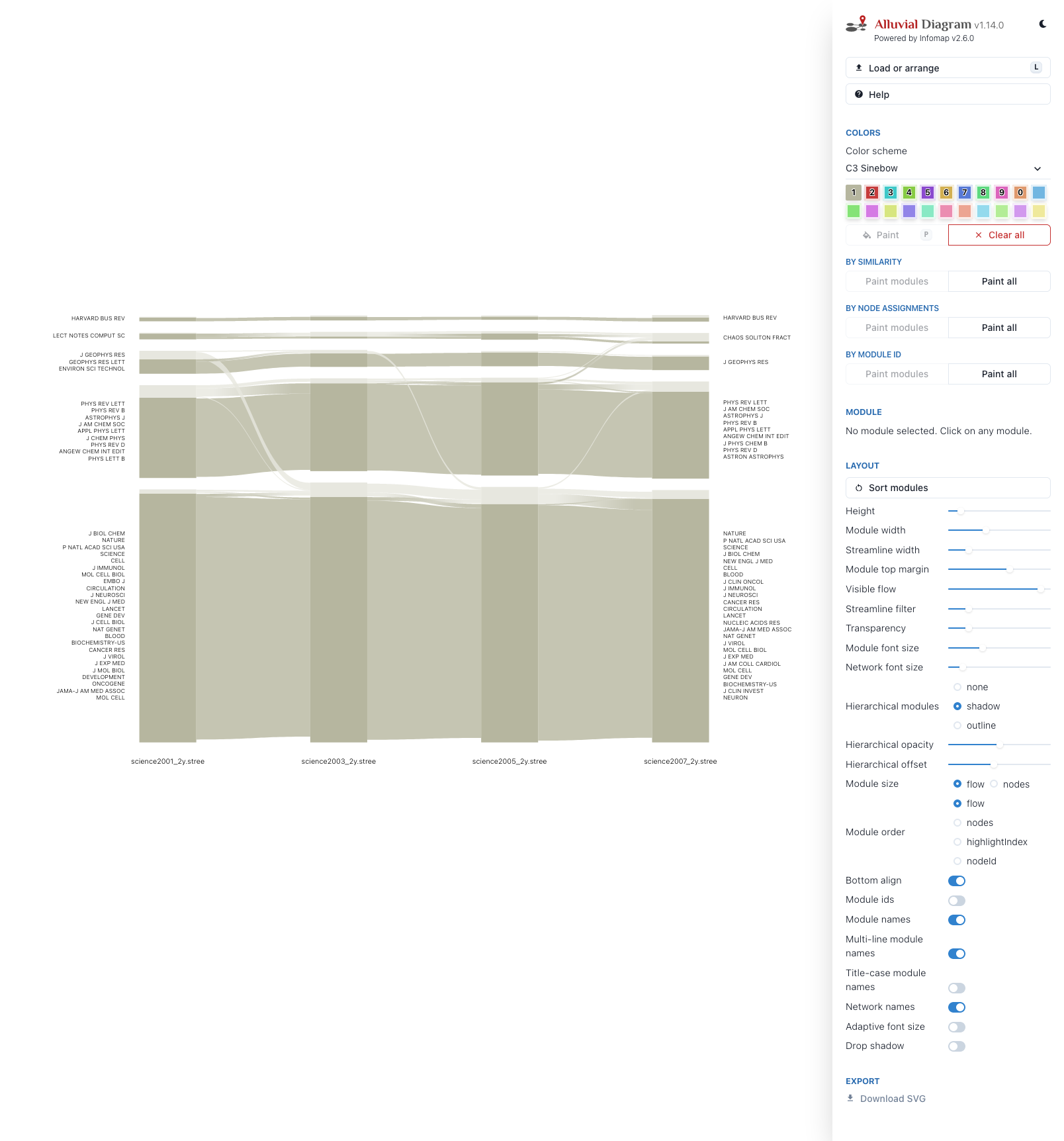}
    \caption{Example session of the alluvial diagram generator.
    The main diagram view can be zoomed in by scrolling in and out. Clicking any empty area and dragging pans the view. 
    Clicking any module (dark rectangles) selects it.
    Double-clicking a module (keyboard shortcut \texttt{E}) expands and shows its submodules if they exist.
    Holding the shift key while double-clicking a submodule (keyboard shortcut \texttt{C}) contracts and regroups the submodules' siblings into their super-module.
    A selected module can be moved around with the \texttt{W} and {S} keys, and networks can be re-ordered left-to-right with the \texttt{A} and \texttt{D} keys.
    The sidebar to the right contains all settings and controls, some available only when a module is selected by clicking on it.
    From top to bottom: ``Load or arrange'' opens the dialogue box in Fig~3.
    ``Help'' displays a help dialogue.
    ``Colors'' allows the user to color all or selected modules with the chosen color scheme.
    ``Module'' contains information about the currently selected module.
    ``Layout'' controls various diagram layout settings, such as size, module order, font sizes, etc.
    Finally, ``Export'' allows the user to export the diagram to an SVG file.}
    \label{fig:screenshot}
\end{figure*}

\begin{figure*}[htp]
    \centering
    \includegraphics[width=\textwidth]{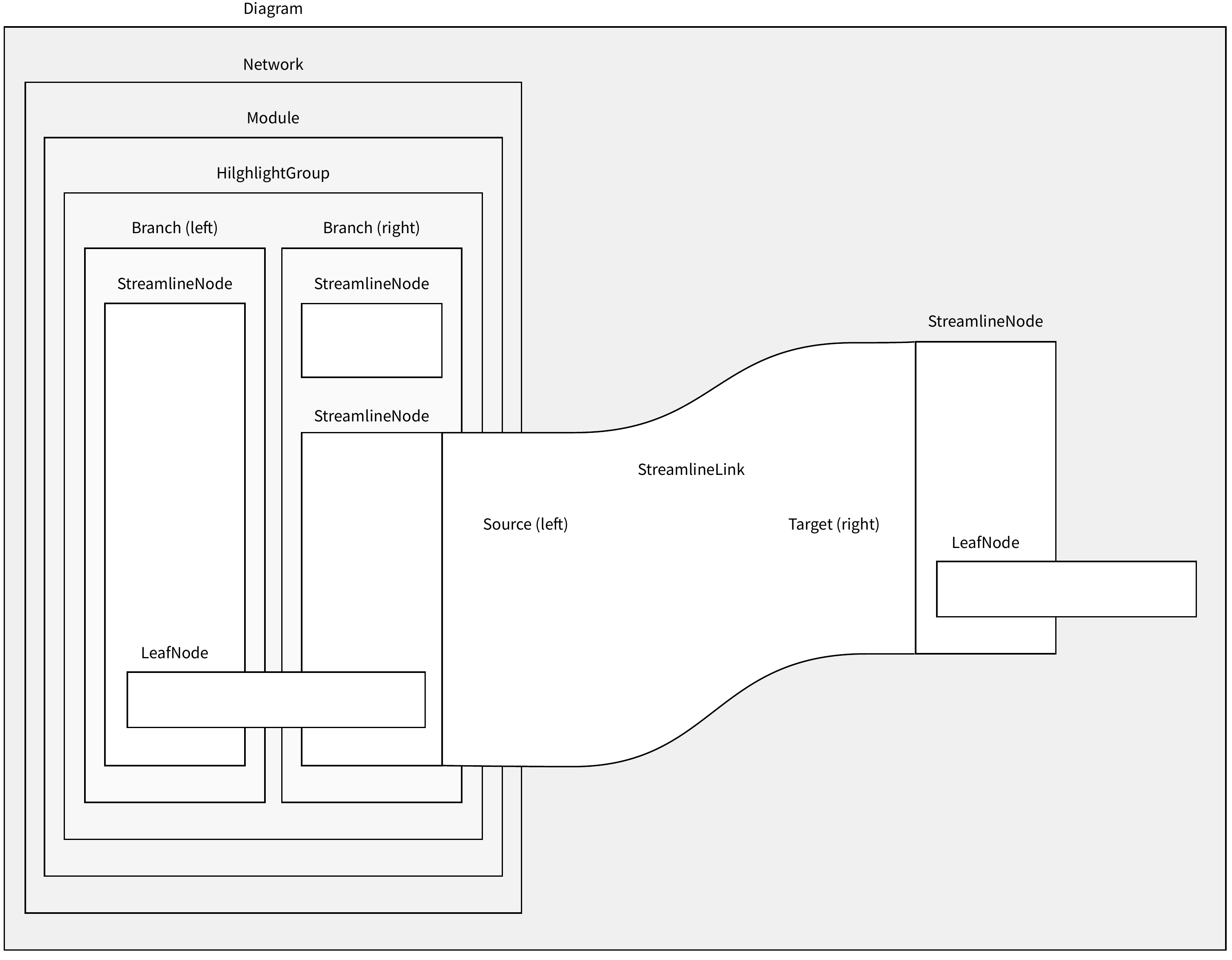}
    \caption{Alluvial diagram object model.
    Each network partition is modeled as a \textsf{Network} whose children are \textsf{Module}s.
    To allow highlighting individual nodes, we divide each module into many \textsf{HighlightGroup}s.
    To connect stream fields between \textsf{HighlightGroup}s in different \textsf{Network}s, each \textsf{HighlightGroup} has two \textsf{Branch}es, \textsf{left} and \textsf{right}, each containing as many \textsf{StreamLineNode}s as there are connected stream fields.
    \textsf{StreamLineNode}s contain many \textsf{LeafNode}s, representing the original nodes in the network.
    Finally, \textsf{StreamLineLink}s connect \textsf{StreamLineNode}s in different \textsf{Network}s.}
    \label{fig:diagram-code}
\end{figure*}

\end{document}